# Cognitive maps and schizophrenia

Matthew M Nour[1,2,*], Yunzhe Liu[3,4], Mohamady El-Gaby[5], Robert A McCutcheon[1], Raymond J Dolan[2,3,6]

1. Department of Psychiatry, University of Oxford, Oxford, OX3 7JX, UK
2. Max Planck University College London Centre for Computational Psychiatry and Ageing Research, London, WC1B 5EH, UK
3. State Key Laboratory of Cognitive Neuroscience and Learning, IDG/McGovern Institute for Brain Research, Beijing Normal University, Beijing, 100875, China
4. Chinese Institute for Brain Research, Beijing, 102206, China
5. Nuffield Department of Clinical Neurosciences. University of Oxford, Oxford, OX3 9DU
6. Wellcome Centre for Human Neuroimaging, University College London, London, WC1N 3AR, UK

* **Correspondence:** Matthew M Nour, matthew.nour@psych.ox.ac.uk

## Abstract
Structured internal representations ("cognitive maps") shape cognition, from imagining the future and counterfactual past, to transferring knowledge to new settings. Our understanding of how such representations are formed and maintained in biological and artificial neural networks has grown enormously. The cognitive mapping hypothesis of schizophrenia extends this enquiry to psychiatry, proposing that diverse symptoms - from delusions to conceptual disorganisation - stem from abnormalities in how the brain forms structured representations. These abnormalities may arise from a confluence of neurophysiological perturbations (excitation-inhibition imbalance, resulting in attractor instability and impaired representational capacity), and/or environmental factors such as early life psychosocial stressors (which impinge on representation learning). This proposal thus links knowledge of neural circuit abnormalities, environmental risk factors, and symptoms.

> *"If the organism carries a 'small-scale model' of external reality and of its own possible actions within its head, it is able to try out various alternatives, conclude which is the best of them, react to future situations before they arise, utilise the knowledge of past events in dealing with the present and future, and in every way to react in a much fuller, safer, and more competent manner to the emergencies which face it"*
>
> <div align="right">Kenneth Craik, The Nature of Explanation, 1943 [1]</div>

## Representation- Biological Psychiatry's missing mediating layer

Psychiatry lacks a mechanistic understanding of how neurobiological abnormalities cause symptoms. This explanatory gap impedes development of theory-guided treatments and has contributed to a decades-long stagnation in clinical outcomes [2,3]. One reason stems from the nature of psychiatric symptoms and signs ("phenomena"), which - to a first approximation - reflect aberrations in cognition and goal-directed behaviour. In schizophrenia, these phenomena include delusions, hallucinations, conceptual disorganisation, and impairments in abstract/inferential reasoning, planning, and social functioning [4–7]. How are we to understand these clinical and cognitive manifestations at the level of the brain?

Psychiatric phenomena reside at a different level of explanation to the neurophysiological processes from which they emerge. Symptoms are *about* things in the world in a way that synapses are not [3,8]. Accordingly, to bridge a divide between neurophysiology and symptoms, brain-based explanations must address how neural activity comes to be *about* things in the world. This necessitates considering the mediating layer of neural **representation** (see **Glossary**) [8].

In this Review we outline advances in cognitive neuroscience that concern how the brain forms structured internal representations of the world - **cognitive maps** - that organize knowledge to guide learning, inference, and behaviour [9–12]. While much of this knowledge concerns hippocampal-entorhinal cortex representations of physical space, growing evidence indicates that conserved algorithmic principles are at play across diverse task domains. We discuss how this programme can inform an understanding of psychiatric phenomena, with a focus on schizophrenia.

## Cognitive maps: the core of cognition and adaptive behaviour

The study of cognitive maps began with a careful observation of behaviour. In the first half of the 20[th] century, psychologists such as Tolman, Harlow, and Craik speculated on the existence of structured internal representations on the basis of observing animal behaviour that was difficult to reconcile with contemporary behaviourist theories of decision making [1,10,13,14]. Tolman described experiments in which rats appeared to learn detailed "field maps" of spatial environments in the absence of reward ("latent learning"), update these maps following surprises, and use them when engaging in deliberative behaviour ("vicarious trial and error") [13]. Harlow showed that – in some cases – primates exhibited task knowledge that could **generalize** beyond the context in which it was acquired, to accelerate learning in new tasks that shared a common underlying structure ("learning to learn") [14].

Here, a deep insight is that that behaviour can strongly imply, and sometimes guarantee, the existence certain kinds of internal representations. This fact constituted a fatal blow to the central tenet of behaviourism - that a discussion of cognitive states is beyond the reach of experimental psychology - and laid the groundwork for the cognitive revolution of the later 20[th] century [15].

In the modern era, our understanding of how internal representations guide behaviour has progressed within diverse mathematical frameworks such as **Reinforcement Learning (RL)**, hierarchical Bayesian inference, and deep neural networks [16–18]. In RL, a particularly influential intellectual strand in cognitive neuroscience, agents are endowed with task representations comprised of states (*"where am I?"*) and actions (*"what can I do here?"*). An agent's task representation, in combination with its learning and decision-making algorithms, profoundly shapes behaviour. This is exemplified by a classical distinction is between "model-free" and "model-based" RL agents.

In RL, model-free agents are endowed with a simple task representation in which state-action values are updated through trial-and-error, akin to the stimulus-response mechanisms of behaviourism (although, even here, representation may not be trivial, see **Box 1**). Model-based agents, by contrast, learn more complex, structured internal models that account for how states *are related*, akin to Tolman's cognitive maps (**Figure 1A**), enabling them to engage in planning, non-local **credit assignment**, and counterfactual reasoning [16]. The model-free vs. model-based distinction speaks to differences in both internal representation and algorithm. In practice, however, these facets are deeply connected: the structure of an agent's task representation has profound implications for the computational efficiency and inferential reach of model-based algorithms [12].

What makes a good task representation? The best task representations capture environmental features that maximally facilitate prediction, inference, and decision-making, and promote re-use of knowledge in new environments [9,10,12,19] *("solving problems in representation, not by exhaustive computation"* [12]). This often necessitates tracking **latent states,** which are inferred as opposed to directly observed (**Box 1**). We note that these representational desiderata are echoed in recent discussions of internal world models in artificial intelligence (AI) and Active Inference [20].

For psychiatry, the key point is that internal representation and behaviour are inextricably linked. Representation can be inferred from behaviour *precisely because* representation constrains behaviour. Clinically-relevant biases in inference and planning arise from task representations that emphasise behaviourally irrelevant environmental features and/or postulate incorrect latent causal structures. To understand how representation-level pathology might arise from neurophysiological dysfunction we first need to understand the neural coding motifs that support cognitive mapping itself.

## Neural mechanisms of map construction

In order to facilitate inference, planning, and knowledge generalisation, the brain must construct internal representations that are structured in a particular way. Starting in the 1970s, electrophysiological studies in behaving rodents have uncovered neural activity patterns in hippocampal formation and prefrontal cortex (PFC) that bear striking correspondence to representational motifs initially conceived through observation of behaviour. Here, we provide a selected overview of key results (for more detailed reviews, see [10,12,21]).

### *Maps of space in hippocampal-entorhinal cortex*

The hippocampal formation encodes environmental states and the relationships between them. During behaviour, hippocampal **place cells** encode the animal's current location in task space (i.e., state), which may be synonymous with spatial location in simple tasks (i.e., place fields) [22–24], or correspond to location in a latent task representation in more complex tasks (e.g., those involving **state aliasing**, see **Box 1**) [12]. Neurons in neighbouring medial entorhinal cortex and subiculum encode relationships between states, **abstracted** from task-specific sensory information ("structural codes", such as **grid cells, object and boundary vector cells**) [25–28] (**Figure 1B**).

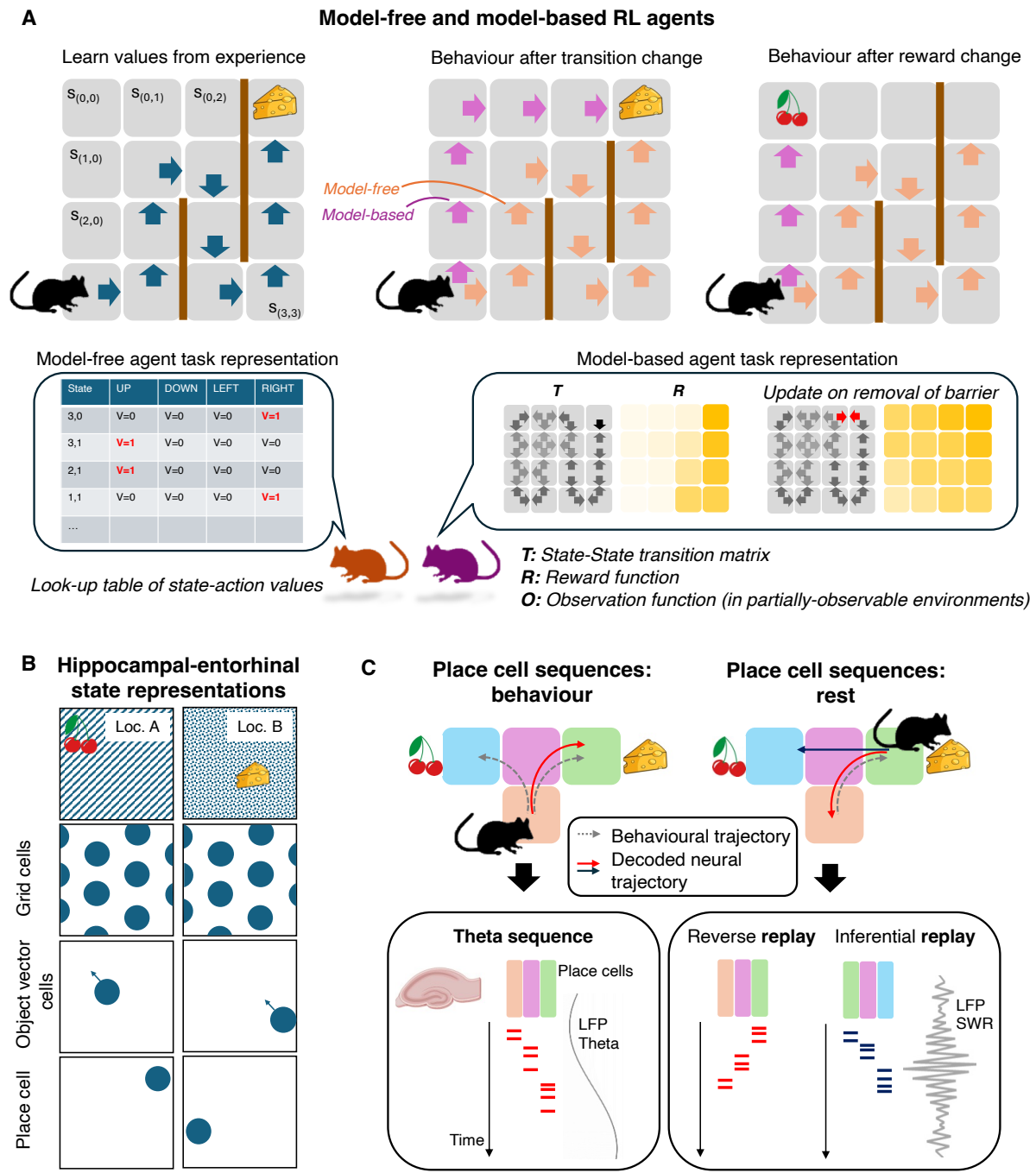

**Figure 1. Cognitive maps: neural and behavioural correlates.**
**(A)** In RL, model-free agents learn the value of each state-action pair through direct experience, and store this value in a form of look-up table. The decision to perform action $A_1$ or $A_2$ at state $S_1$ is informed by how likely each state-action pairing was to lead to reward in the past (i.e., $V(A_1, S_1)$ vs. $V(A_2, S_1)$). (Temporal discounting of value estimates has been omitted.) By contrast, model-based agents learn an internal model of the task structure, comprising action-dependent Markovian state transition functions ($P(S_2|S_1, A_1)$), reward functions ($P(R|S_1, A_1)$), and observation functions ($P(O_1|S_1)$). Model-based agents can update policies when the environment changes (barriers removed, new rewards added).
**(B)** Selected spatial representations in hippocampal-entorhinal cortex (adapted from [12]), spanning medial entorhinal structural codes (grid cells and object vector cells [25–28]) and hippocampal place cells [22–24].
**(C)** Place cell sequences betray spatial map computation. T-maze, where rodents are trained to turn left or right at a junction (never running the full top segment). *Left*: "Look-ahead" theta sequences during behaviour (supporting spatial memory and planning), during local field potential (LFP) theta oscillations [21,29–32]; *Right:* Replay sequences during rest (reverse replay, supporting credit assignment [34], and sequences that "stich together" experiences, supporting relational inference [40,41]) during LFP sharp wave ripple (SWR) oscillations.

Hippocampus also encodes structured information about task states in cell assembly sequences. During behaviour, place cell **theta sequences** chart future behavioural trajectories in a manner that supports sequential memory and planning [21,29–32], while, during resting periods, hippocampal **replay sequences** recapitulate past experiences in both forward and backward directions in a manner thought to support memory consolidation and credit assignment [33–36] (**Figure 1C**).

Hippocampal assembly sequences also encode information that goes beyond direct experience. Thus, replay can play out novel behavioural trajectories [37–39], "stich together" fragments of experience [40,41] (**Figure 1C**), and display non-behavioural sampling statistics [42]. **Preplay** sequences have also been observed, so-called because the episodes they encode do not appear to correspond to any prior environment, but instead acquire behavioural relevance with respect to place maps learned in future environments (potentially reflecting a role in scaffolding new experiences) [43,44].

Together, these observations constitute strong evidence that hippocampus is key to organising knowledge of how environmental states are structured. They also point to a role in knowledge generalisation, whereby as-yet unexperienced associations (state-state relationships or conjunctions) are inferred through an application of prior knowledge about the relational structure of the environment. Indeed, it has long been known that lesions to hippocampal formation in rodents impair relational inference [45,46].

*Generalisation through abstraction*

The computational principles underlying knowledge generalisation are incompletely understood. Leading accounts implicate structural codes, as carried in grid and object vector cells. These codes contain information about the way states are related, where such information is abstracted from any specific sensory experience, conserved across contexts, and factorized (**Box 1**). This enables structural codes to act as generalisable basis set – a repertoire of representational "building blocks" - for new map construction. These building blocks (or primitives) can be flexibly combined to form conjunctive state representations in new environments [9,10,12,17,44,47,48], potentially implemented in hippocampal replay [9,12,44,48]. Compositional coding provides a powerful mechanism for knowledge generalisation: primitives come pre-packaged with structural information that permits zero-shot relational inferences (*"a barrier separates location A and B"*) and behavioural competencies (*"avoid barriers"*) [9,12].

Thus, knowledge generalisation and state space construction can be cast as inference on the compositional structure of the environment [12], in which the repertoire of representational building blocks – potentially carried by hippocampal-entorhinal structural codes – constitutes prior information about the latent structure and behavioural affordances of new environments. Under this account, pathologies of generalization – such as those seen in situational anxiety and paranoia - might stem from abnormal inference itself (weighting of prior and likelihood information), or from biases in the composition and distribution of one's prior structural repertoire (as discussed below). This constitutes an under-explored point of contact between cognitive mapping and Active Inference/Predictive Coding accounts of schizophrenia, and psychiatric syndromes more broadly (**Box 2**).

## Domain general cognitive mapping

For cognitive mapping to serve as a viable explanatory framework in psychiatry, the aforementioned neural coding schemes must extend beyond a representation of physical space. Indeed, as early as the 1970s O'Keefe and Nadel speculated that hippocampus might support domain general cognitive mapping [23] – a proposal now supported by abundant evidence.

Animal studies report place- and grid-like coding of olfactory [49], auditory [50], and social [51] variables, as well as in tasks involving non-spatial relational inference [41]. Human functional neuroimaging studies, in combination with innovative analytic techniques (**Box 3**)**,** also provide evidence for domain-general cognitive mapping in hippocampal formation. Thus, using functional magnetic resonance imaging (fMRI), the **representational geometry** of hippocampal-entorhinal cortex task-evoked activation patterns has been shown to mirror latent task structure in an abstracted manner that facilitates knowledge generalisation [52,53] [54]. fMRI has also been used to uncover grid-like coding in entorhinal cortex during spatial navigation (virtual [55] and imagined [56]), and when participants "navigate" structured conceptual and social spaces [57,58].

fMRI and magnetoencephalography (MEG) have also been used to index spontaneous sequential neural state reactivations, indicative of replay, in tasks involving sequential inference and decision-making [48,59,60]. Such activity bears several correspondences to hippocampal replay in rodents. fMRI replay localises to hippocampus [59], while, in MEG, replay reflects inferred (as opposed to merely experienced) task structure, and is coincident with high-frequency oscillations that source-localise to hippocampus (akin to hippocampal **sharp wave ripples** (SWRs), **Figure 1C**), and reverses in direction following reward [48]. MEG-measured replay is also composed of factorised representations of abstracted structural codes and sensory codes (see **Box 1**), where the former can be detected prior to task exposure and may facilitate structural knowledge transfer (generalisation) [48] (cf. rodent preplay [43,44]). Remarkably, similar non-spatial hippocampal-entorhinal codes have recently been identified in human intracranial recordings, including experience-dependent changes in representational geometry, state tuning, and sequential replay [61].

Beyond medial temporal lobe, human medial PFC (mPFC) also exhibits hallmarks of domain-general cognitive mapping, including grid-like coding [55,57,58], abstracted task schemas [62] and conjunctive state representations that reflect knowledge of task structure [47,63]. In rodents, PFC tracks progress to internally-generated goals, reflecting use of cognitive map representations in flexible planning [64]. Likewise, orbitofrontal cortex is causally implicated in representation of latent states, as revealed by tasks involving state aliasing [59,65–67]. The latter findings raise an intriguing hypothesis that PFC might encode the "structure of an agent's goal-directed behaviour", which can be contrasted to hippocampal formation's specialisation in encoding the "structure of the world" [64].

An exciting research area concerns how different brain regions interact to support cognitive mapping functions. Goal-centric PFC representations also encode spatial information [64] (presumed to originate in hippocampal-entorhinal cortex), while hippocampal place cells are biased by goal information [68,69] (presumed to originate in PFC). In rodents, non-human primates, and humans hippocampal replay and SWRs are temporally correlated to neocortical activity fluctuations [70–74]. Here, there appears to be some specificity for default mode network (DMN) [72–74], a collection of predominantly-midline cortical regions considered to occupy the deepest layers of a cortical processing hierarchy and proposed to support world models at the highest level of abstraction (e.g., narrative schemas) [75–78].

## Cognitive mapping and schizophrenia – behavioural and neural evidence

Before moving to a discussion of how a cognitive mapping framework might inform psychiatry, we offer a brief distillation of our discussion thus far. First, an understanding of how internal representations are acquired and structured is key to understanding the diversity of behaviour and cognition, including the extremes that are the focus of psychiatrists and clinical psychologists. Second, a great deal about the structure of internal representations, including candidate neural coding motifs, can be inferred from a careful study of behaviour, particularly in conjunction with computational modelling. Third, the recent focus on neural map representations in cognitive neuroscience has confirmed the existence of many such coding motifs. This research programme is now sufficiently mature to begin to inform a new era of mechanistic models in psychiatry.

In the remainder of this Review, we turn to the specific instance of schizophrenia. We begin by considering how a cognitive mapping account might accommodate clinical features in people with a diagnosis of schizophrenia (PScz), and highlight behavioural and neural evidence for cognitive map dysfunction. We then speculate on how map dysfunction might arise from potential upstream causal factors, focusing first on neurophysiological abnormalities, and second on (somewhat independent) early-life environmental risk factors.

*Starting with behaviour*

Many symptoms and signs in schizophrenia can be understood as manifestations of abnormalities at the level of internal representation, broadly dichotomised as those affecting inference and generalisation (e.g., delusions, paranoia, ideas of reference), and those affecting sequential sampling from underlying representations (e.g., formal thought disorder) (**Table 1**).

In behavioural experiments using curated tasks, PScz demonstrate impairments in model-based decision-making [79,80], relational inference [81–84], and sequential planning [5–7,85]. A cognitive mapping framework additionally invites us to consider behavioural consequences in more abstract task domains, where the state space comprises concepts, and relational structure corresponds to pairwise semantic similarity. Approaches to operationalising this more abstract relational structure include inferring it from participant behaviour (e.g., word association data [86] or similarity judgements [87]) or using learned embedding spaces from pretrained AI language models. Intriguingly, there is tentative evidence that some aspects of these "semantic spaces" are represented by coding motifs originally evolved for spatial navigation, in line with domain-general cognitive mapping theories. Thus, behaviour in category fluency tasks (*"name as many animals as you can"*) is well described by patch foraging models of animal search behaviour [88], and is accompanied by hippocampal activity patterns that track semantic category switches [87] and distances [89].

In PScz, language-based studies report abnormalities in semantic distances separating consecutive words and utterances [90]. In one study employing a category fluency task, word lists generated by PScz were significantly less predictable using semantic relatedness information derived from an AI language model (**Figure 2B**) [91], reminiscent of early accounts of schizophrenia as a disorder of "loosened associations" (**Table 1**).

*Neural evidence of cognitive map dysfunction*

Neural signatures of replay and grid coding have recently been investigated in PScz using functional neuroimaging during tasks that require participants to use knowledge of how states are related.

In one MEG study participants were tasked with inferring how pictures were sequentially related, by combining information gathered from direct experience and a pre-learned abstracted task schema (i.e., a structural prior) [84]. PScz displayed impaired neural replay for inferred task structure and abnormal replay-related hippocampal ripple oscillations in the rest period immediately following learning, despite no behavioural evidence for impaired ceiling-level knowledge or accelerated forgetting (**Figure 2C**), reminiscent of genetic mouse models of schizophrenia [68,92,93] (discussed below). There were also differences between PScz and control participants in the representational geometry of task-evoked activations, with control participants alone displaying an abstracted code for ordinal position that emerged as a function of learning.

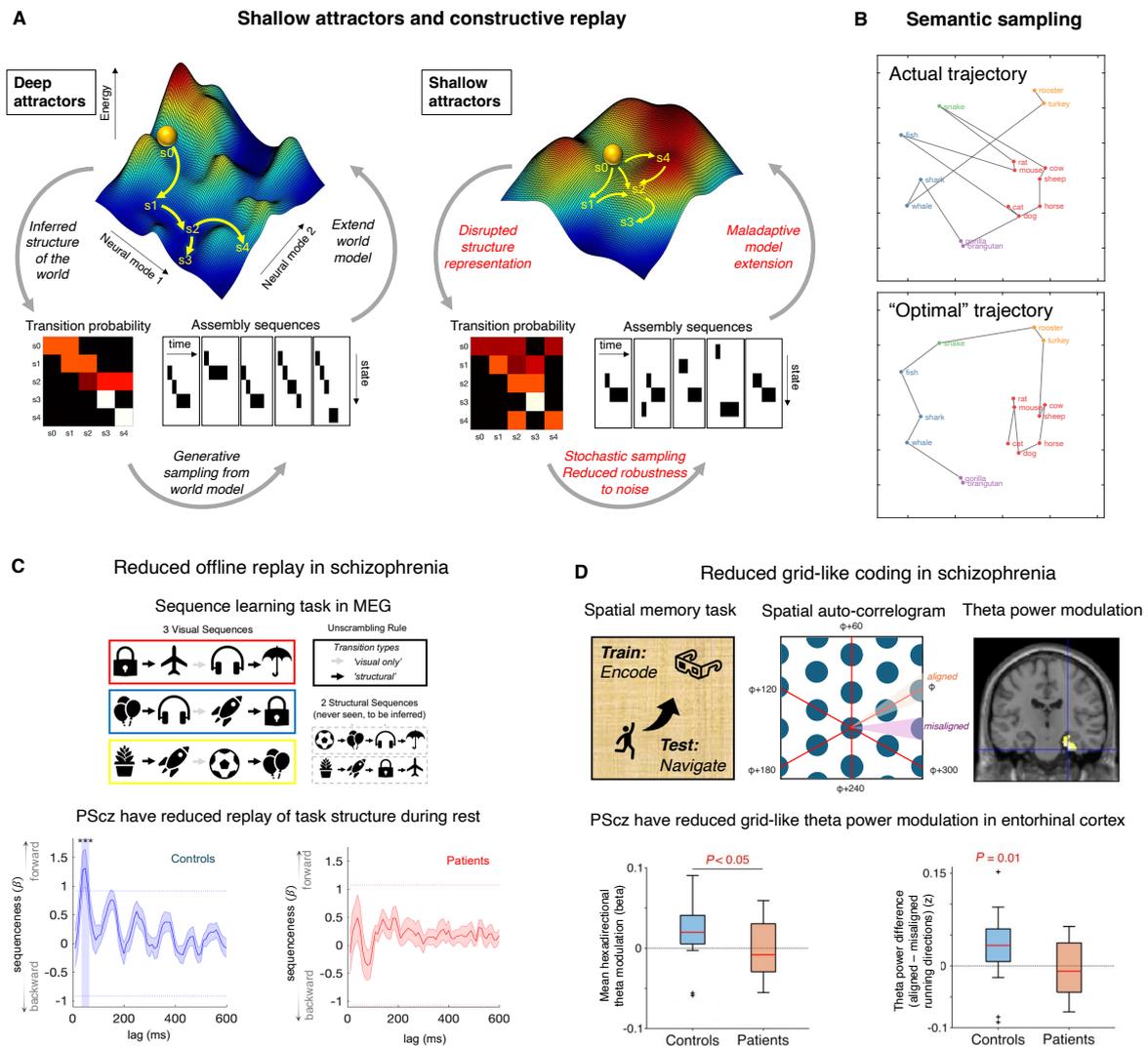

**Figure 2. Cognitive mapping and schizophrenia**
**(A)** In hippocampus, shallow attractors [98,110] predispose to abnormal sequential replay, potentially leading to further entrenchment of abnormal attractor basins (see main text). Schematic inspired by [110].
**(B)** In a category fluency task, words can be represented as high-dimensional vectors in semantic space using an AI word embedding model. Each participant's word list can then be construed as a trajectory through semantic space (here a 2-dimensional projection of a 300-dimensional embedding). The "actual" trajectory can be compared to one that minimises semantic distance "travelled" ("optimal"). At the group level, PScz word lists deviated from "optimality" more than controls [91]. Figures from [91].
**(C)** *Top:* MEG sequence learning task. Participants need to infer the correct transitions between 8 task pictures ("structural sequences") [84]. *Bottom:* During a post-learning rest session, PScz exhibited reduced evidence for spontaneous neural replay of task structure compared to controls (using TDLM, see **Box 3**). In controls, replay was evident at ~50ms state-state transition lag. Figures from [84].
**(D)** *Top left:* MEG spatial memory task. Participants navigate to remembered locations in a virtual arena. *Top middle:* Exemplar grid cell spatial auto-correlogram. In functional neuroimaging studies, grid-like coding can be indexed by modelling neural activity as a function of movement direction (e.g., grid-aligned vs misaligned, see **Box 3**) (schematic inspired by [55]). *Top right:* Grid-like theta modulation in right entorhinal cortex across all participants. *Bottom:* Grid-like coding was reduced in PScz (data figures from [84]).

In subsequent work, the authors identified impaired replay-DMN coupling in PScz [94], and – using positron emission tomography in the same participants - an association between replay-associated ripple power and hippocampal NMDA-receptor availability [95] (thought to play a crucial role in cortical E/I balance and replay [96–98]).

In another MEG study using a spatial memory task, PScz displayed reduced theta phase coupling (1-8 Hz) between mPFC and medial temporal lobe during cued spatial memory recall [82], in line with findings in two animal models [99,100]. PScz also exhibited reduced entorhinal grid-like coding (4-10 Hz power modulation) during virtual spatial navigation [101] (**Figure 2D**).

These clinical MEG studies thus provide the first direct evidence for impaired hippocampal-entorhinal cognitive mapping signatures in PScz.

## Paths to symptoms: cognitive maps as a mediating layer

How might cognitive map dysfunction arise from upstream neuropathological and environmental risk factors?

### *Neurobiology: from shallow attractors to symptoms*

Schizophrenia is associated with diverse neurobiological abnormalities in brain regions that support cognitive mapping. In hippocampus, this includes reduced grey matter volume [102], reduced synaptic density [103], increased resting state perfusion and metabolism, and abnormal task-related activation [104]. Similar abnormalities are reported in prefrontal cortex [102,103,105,106]. However, such findings are rarely anatomically-specific and do not shed light on how neurophysiological disturbances result in representation-level abnormalities that drive specific symptoms.

More promising, are convergent findings that a final common pathway in the neurobiology of schizophrenia is a disruption in the balance between excitatory glutamatergic and inhibitory GABAergic signalling in cortical circuits, termed excitation-inhibition (E/I) imbalance, as evidenced by electroencephalographic abnormalities in gamma oscillations and sensory gating deficits (among many other findings) [97,98,107].

From a dynamical systems perspective, E/I balance in recurrent cortical circuits underpins **neural attractor** dynamics – a network's ability to maintain stable population activity patterns [98,108–110]. By contrast, E/I imbalance disrupts a neural attractor landscape, rendering attractors unstable or "shallow" – that is, increasingly sensitive to internal or external noise [110]. Consistent with this account, genetic rodent models of schizophrenia exhibit a reduction in the number of stable cortical activity states [108,109].

The ability of a network to maintain stable attractors underpins both its representational and computational capacities. This property makes explanations couched in terms of abnormal attractor dynamics a promising candidate for explanatory models in psychiatry. Accordingly, such explanations have been advanced to explain cognitive and behavioural symptoms not only in schizophrenia, but also in other conditions involving working memory impairment, inattention and behavioural variability, such as attention deficit hyperactivity disorder (ADHD) [98,109,111,112].

A cognitive mapping perspective casts neural attractor instability in a subtly new light, considering not only its immediate consequences (e.g., sensitivity to noise), but also downstream effects on the structure of an agent's internal representations [110]. This is because many of the discussed hippocampal-entorhinal neural coding motifs – from spatially-tuned place and grid fields, to sequential replay – can be construed as neural attractors [110].

In PScz, the evidence for attractor instability in hippocampal-entorhinal codes is indirect, and includes the aforementioned MEG studies. However, there is abundant evidence for hippocampal place coding disruption in rodent models of schizophrenia, which are engineered to reflect known genetic and neurodevelopmental risk factors (e.g., *22q11.2* deletion / *Setd1a* mutation [68,109], and maternal immune activation [113], respectively; see [114] for an overview). Here, notable findings include abnormal elevations in population-level SWR power and rate [68,92,93,113], disrupted place cell coactivation patterns during rest [92] (a prerequisite for replay), and disrupted place cell theta sequences, theta phase coupling, theta phase precession, and gamma-theta coherence [113,115]. Animal studies also find population place map instability over time, impaired goal-related remapping [68], and overly generalized place field coding [116].

Thus, E/I imbalance, as seen in (rodent models of) schizophrenia, has a potential to disrupt hippocampal (and presumably, entorhinal) coding motifs that have been directly tied to neural state representation and cognitive map construction. Importantly, such representations need to be maintained in the absence of driving sensory information, either because they manifest during rest periods (i.e., replay), or because the mapping between on-task sensory information and corresponding latent state representation is indirect and weak (**Box 1**). It is in precisely such situations that attractor instability is most vulnerable to being unmasked [110].

One speculative path from attractor instability to symptoms concerns the role of replay in cognitive map construction. During rest, population activity governed by shallow attractor dynamics may transition between states that are "far apart" in representational space, or flit between neighbouring attractor basins. This results in novel ensemble sequences and compositions [44] that are implausible under an appropriately conditioned world model. Such information might nevertheless be used to update and extend map representations, entrenching abnormal attractor basins, which in turn constrain further cycles of inference, generative/constructive replay, and map extension (**Figure 2A**). Speculatively, in clinical settings these abnormal dynamics might manifest as conceptual disorganisation and belief instability (flitting between weakly associated conceptual states – a direct consequence of shallow attractors), or delusions (entrenched, yet inappropriate, attractor basis – arising from the interaction between shallow attractors and constructive replay) (**Table 1**) [98,110,112]. This dynamical account dovetails with clinical observations that delusions may initially be somewhat malleable, before crystalising into more incorrigible, fixed beliefs (see also **Box 2** and [117]).

Another path to symptoms concerns task domains that necessitate maintenance of latent state representations, which, by definition, are not resolvable from sensory information (a driving stimulus) alone, and are thus vulnerable to neural attractor instability. Latent state representations are indispensable in tasks that require abstract reasoning (i.e., cognition on structural features), planning (requiring hierarchical task decomposition into subgoals), social cognition (where latent variables span social kinship graphs, role-relationship schemas, and the mental states of others), and context-dependent behavioural routines (necessitating formation and maintenance of overarching goals and task schemas) (**Box 1**). An impaired ability to maintain stable latent state representations might account for some aspects of cognitive impairments in schizophrenia, particularly in social cognition and executive functions spanning abstract reasoning, goal maintenance, and planning [5–7,85]. More speculatively, this impairment may also relate to delusions, many of which comprise beliefs about the latent structure of the world (e.g., webs of interactions between people or entities in paranoia), and one's location within it (e.g., the centre of the web).

*The early environment: from biased representation learning to symptoms*

Any viable model of schizophrenia – or indeed, any psychiatric condition – must be capable of accommodating the considerable influence of vulnerability factors such as carer abuse, parental loss, bullying, immigrant status [4]. How are we to understand the role of these environmental factors within a brain-based model of schizophrenia?

Previous attempts at an integrated model have adopted one of two avenues. First, is to note that early life risk factors – including psychosocial stress - exert lasting effects on neurobiology (e.g., synaptic spine density and dopaminergic signalling [4,105]), likely to impact neural processes important for neural computation (e.g., E/I balance). Second, is to posit that psychosocial stressors foment maladaptive cognitive schemata that bias appraisals, potentially leading to symptoms such as paranoia [4]. It has been difficult to integrate both accounts within a single mechanistic model of neurocognitive function.

As discussed, under a cognitive mapping framework, state space construction is cast as inference over compositional structure [12]; agents do not construct representations of new environments *de novo*, but instead re-use (compose) representational building blocks (primitives) from a pre-existing repertoire [9,10,12]. Compositional primitives come pre-packaged with information about how states are related (i.e., abstracted structure) and state-action values that bias behaviour (i.e., pre-credit assigned) [9,12], and may themselves be composed of yet simpler primitives [12,44,118,119] (perhaps discovered through replay [9,44]). Thus, an agent's task representation (posterior distribution over structure) will be exquisitely sensitive to both the content and probability weighting of their repertoire of primitives (priors) (see **Box 2** for connections to hierarchical Bayesian accounts of psychopathology).

Recent work on representation learning in artificial neural networks confirms that this representational repertoire is influenced by the model's training data. Within meta-RL, the distribution of tasks an agent encounters shapes how it "learns to learn" in new domains – including heuristics on environmental reward rates, controllability, and volatility [120]. Relatedly, in disentangled representation learning, if an agent encounters a training environment in which two (actually distinct) latent variables are spuriously correlated, it will be impossible to learn appropriately disentangled neural representations of the environment's true latent causal structure (**Box 1**), leading to negative transfer effects ("*entangled tasks lead to entangled representations*" [12]). Finally, recent work in hierarchical concept learning demonstrates that the order in which information is presented during learning has profound consequences for how learning unfolds, and that it is easy to get stuck in maladaptive learning traps when faced with chaotic curricula [118].

The connection to psychiatry is that model training data might serve as a useful abstraction of an early developmental environment [120]. Indeed, multiple known early environmental risk factors for adult psychiatric conditions – such inconsistent caregiver behaviour, poverty, pervasive bullying, and the reliable co-occurrence of authority figures and abuse - can be cast as disruptions in the content or statistics of this critical early environment. We speculate that such biases might promote learning of structural heuristics and primitives (schemata) that predispose to maladaptive inferences in adult life (helplessness, paranoia), that in turn manifest as psychiatric syndromes. It is plausible that this process might occur independent of any "abnormality" at the level of neurophysiology.

The claim that early life experience biases cognitive schemata is not new [4]. Yet, the cognitive mapping framework – by placing the question of neural representation centre stage – presents a unique avenue for theorising about, and testing, the effects of specific stressors on representational structure. One starting point might be to use "toy model" artificial neural networks that permit in silico interventions pertaining to both environmental (training data) and biological risk factors (learning rules and network architecture) [118,121,122].

## Concluding remarks

We began by reflecting that biological psychiatry needs to advance models that are capable of explaining how neurophysiological perturbations and environmental risk factors cause symptoms. We have argued that this is only possible by engaging with a mediating layer of neurocognitive representation, which considers how population neural activity and environmental statistics give rise to structured internal representations of the world (cognitive maps). Cognitive maps influence all aspects of cognition, and their study is now a major focus across neuroscience and AI. We believe that biological psychiatry has much to gain from incorporating these advances, both in understanding disorders such as schizophrenia as well as a myriad other psychiatric conditions.

We have reviewed growing behavioural and neural evidence implicating disorganised cognitive maps in schizophrenia, and have outlined a roadmap for future work that aims at an integrated explanatory model of the condition, spanning an understanding of circuit-level algorithmic disruption and the role of the early-life environment alike.

The ultimate goal is that improved mechanistic understanding will drive improvements in clinical outcomes. This might come from development of biological interventions that restore representational capacity to disrupted neural networks, or psychological interventions that promote learning better structural primitives. We need to acknowledge that this is likely to be a long road and there are many **Outstanding Questions**. Yet – as our discussion illustrates – if we seek to reliably predict and intervene on a system, then we must first understand it.

**Glossary**

**Abstraction:** Distilling information from one domain that is shared with related domains (typically, common structure, "abstracting away" sensory features, referred to as a "structural code").

**Attractor**: Recurrent points or paths (stable states) in neural activity space that emerge from the biophysical and connectivity properties of a recurrent network.

**Cognitive map:** structured internal representations that capture information about the states in an environment and the manner in which they are related. Applies to spatial and non-spatial domains alike.

**Credit assignment:** Updating the value of states and actions following rewards or punishments, according to an internal world model (cognitive map).

**Generalisation (transfer):** The transfer of (suitably abstracted) information from one domain to another.

**Grid cells:** Neurons in medial entorhinal cortex and mPFC that respond when an animal is in one of a set of spatial locations structured in a regular grid tiling a state space. The population grid code is conserved across environments, potentially serving as a global coordinate system (structural code) for generalisation

**Latent (hidden) state:** An inferred environmental state that is not directly observable. A grouping of sensory states based on behaviourally-relevant commonalities.

**Object and boundary vector cells**: Found in medial entorhinal cortex and subiculum, respectively, these neurons encode vectors to salient objects or boundaries, preserving this relational coding across environments (like grid cells). Serve as a local coordinate system (structural code) for generalisation.

**Place cells:** Hippocampal neurons that respond when an animal is in a particular location (state). The population place code is not conserved (i.e., remaps) across environments.

**Preplay**: Hippocampal networks are preconfigured into sequential activation motifs which, in subsequently encountered environments, may come to correspond to behaviourally-meaningful sequences.

**Reinforcement Learning (RL):** Mathematical formalism describing how agents select actions in environments to maximise the sum of (temporally discounted) expected future reward.

**Replay sequences:** During rest (awake immobility and non-REM sleep), place cells fire in time-compressed sequences in conjunction with SWR events.

**Representation:** A neural activity pattern that is "about" a state of the world, can be reinstated in the absence of the eliciting state, and plays a causal role in generating behaviour. Cognition is computation (transformations) over representations [8]. A representation is a functioning isomorphism between a set of processes in the brain, and a behaviourally-important aspect of the world [15].

**Representational geometry:** Similarity structure relating neural representations of task states (**Box 3**).

**Sharp wave ripple (SWR):** Large amplitude deviations coupled with 140-200 Hz "ripple" oscillations seen during rest in hippocampal local field potential (LFP). Accompany replay.

**State aliasing:** Near-identical sensory states that map to distinct latent states.

**Theta sequences**: During behaviour, place cells fire in time-compressed sequence nested within LFP theta cycles, tracing potential behavioural paths.

**Table 1 – Symptoms and signs of schizophrenia through a cognitive mapping lens**
Accompanied by item code from the Positive and Negative Symptoms of Schizophrenia (PANSS) scale [123].

| Symptom or sign | Description | Cognitive mapping perspective |
| --- | --- | --- |
| Delusions (P1, P5, P6, G9) | Fixed, abnormal beliefs, often persecutory or bizarre. | Delusions reflect maladaptive internal representations of environmental states and relational structures, arising from abnormalities in structural generalization and inference. Neurally, these deficits may stem from E/I imbalances, leading to shallow attractor dynamics and disrupted generative replay (see main text). |
| Conceptual disorganisation (P2) | In formal thought disorder, disrupted semantic relationships between successive speech utterances (e.g., "loosening of associations"). | This may reflect disorganization in internal representations (disrupted cognitive maps of semantic space) or in stochastic sampling processes acting on these maps (e.g., consequent upon shallow attractor dynamics, **Figure 2A**). |
| Perceptual abnormalities (P3) | Hallucinations (false perceptions) or illusions (distorted perceptions). | Predictive coding accounts propose that perceptual abnormalities result from disruption at the level of prior beliefs (internal representations) about the causal structure of the environment (e.g., reduced precision) (**Box 2**). |
| Negative symptoms (N1-N4) | Including apathy, amotivation, emotional blunting. | Some negative symptoms might stem from impairments in representing latent task states and goals, and tracking proximity and progress to goals [124] (**Box 1**). mPFC cognitive maps are sensitive to goal information (e.g., progress-to-goal representations [64]). |
| Cognitive impairment (N5, N7) | Deficits in relational inference, sequential planning, and higher-order (e.g., analogical) reasoning [5–7,79–85]. | Planning requires an internal representation of how of states in the environment are related (e.g., action-dependent state-state transition matrix, **Figure 1A**). Relational inference rests on transfer of prior structural knowledge to new setting, enabling inference of "missing link" associations that have not been directly observed. Analogical reasoning rests on an ability to infer a shared relational structure (isomorphism) between two superficially-different domains. |
| Phenomenology (of delusions) | Some patients experience delusions as a radical restructuring of the experienced world, involving internal struggles for understanding and control, and reduced feelings of uncertainty [125]. | These descriptions resemble insight or "ah-ha!" moments: the sudden discovery of a new way of understanding a problem in terms of a familiar and parsimonious latent causal structure - a form of structural knowledge generalisation [118]. Anecdotally, insight usually arises following a period of offline cognition (cf. generative/constructive offline replay [9,12,44]). |

**Box 1: Challenges of map building in the real world**

### Curses of Dimensionality and Partial Observability

Real world experiences are not annotated with "state" and "action" labels. When agents attempt to model the world based solely on sensory data, they face two core challenges. First, each experience contains an overwhelming number of sensory features, creating an expansive state space that is difficult to navigate or learn from (the "curse of dimensionality"). Second, many factors critical for decision-making are not directly tied to sensory inputs ("partial observability") [19].

To overcome these challenges, agents must infer the latent (abstracted) states that constitute an environment. This might involve collapsing across superficially distinct sensory states that are behaviourally equivalent, or conversely, mapping superficially similar sensory episodes to different latent states based on experience history (state aliasing) [19]. The resulting state space is a substrate for learning and decision making algorithms (the states and actions that both model-free and model-based RL algorithms operate on belong to the internal representation, not the world *per se*) (see **Figure 1A** and below).

### Finding good abstractions

Good task representations rely on latent state spaces that are low-dimensional, sparse, and disentangled (or factorized). This reflects a heuristic that the causal structure of an environment is often well approximated by a small set of latent causal factors that act (somewhat) independently, and can thus be understood in isolation. Such representations significantly simplify learning and planning. They also afford generalisation by allowing flexible reuse of knowledge across different contexts [9,10,12,48].

The brain is endowed with inductive biases that afford useful abstractions. These biases are shaped through learning processes acting over both evolutionary and developmental time (e.g., meta-learning and hierarchical concept learning) [10,12,18,118,120].

### The construction of reward

Real world experience also does not come labelled with exogenous "rewards". These, too, must be constructed. Hedonic responses might be a function of an agent's perceived progress towards internally-defined goal states, reminiscent of path integration on an internal representation that spans multiple goal-relevant dimensions (social, financial, etc) [12].

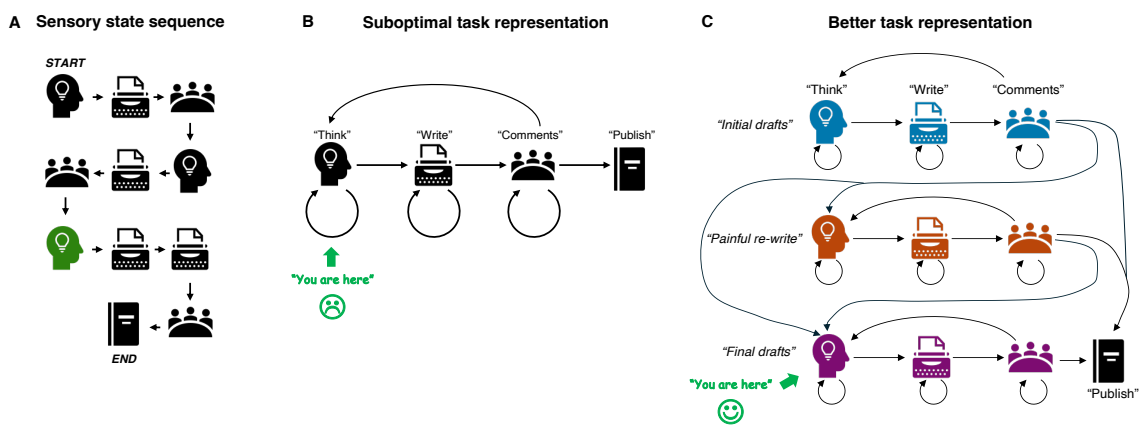

**Navigating and planning in task space**. (**A**) Internal representations of task states and their (action-dependent) relationships allow agents to track progress to goals, interpret new sensory data, and plan. The same sequence of observations (sensory states) can give rise to multiple internal task representations (e.g., **B** and **C**, the latter accounting for state aliasing in a manner that affords improved inference and planning). Inspired by [124].

**Box 2: Cognitive Mapping and Predictive Coding**

### Predictive Coding

Predictive Coding conceptualises the brain as a hierarchical Bayesian inference machine that seeks to maximise the evidence for its internal model of the world. This objective can be cast as minimising the discrepancy (prediction error) between the brain's predictions of observable sensory data, and sensory data itself. Predictions arise from the brain's generative model, and project from higher to lower hierarchical layers (top-down). Sensory data is carried from lower to higher hierarchical layers (bottom-up). The integration between these sources of information can be formalised as Bayesian inference: prior beliefs (predictions) combine with likelihood information (sensory data) to update a posterior belief (new representation of the environment). The magnitude of the belief update is determined by the relative precision (reliability) of prior and likelihood information [126].

Active Inference extends this framework to action. In order to minimise the discrepancy between internal predictions and sensory data, agents can modify which sensory information is sampled (action), in addition to updating internal beliefs (perception) [20].

### Predictive Coding accounts of psychosis

Predictive Coding accounts of psychosis posit that positive psychotic symptoms are a product of aberrant hierarchical inference, resulting in abnormal posterior beliefs about the latent causes generating observable data. This computational abnormality might relate to abnormalities in cortical glutamatergic/GABAergic signalling (that carry prior and likelihood information), and dopamine/acetylcholine signalling (thought to modulate precision-weighting). The canonical predictive coding account posits a reduction in precision of top-down priors (consequent on glutamatergic and GABAergic impairments in frontal cortex) and a strengthening of precision in bottom-up likelihoods (consequent on mesostriatal hyperdopaminergia), resulting in delusions and hallucinations (both construed as "false inferences") [126]. Extensions to this canonical account exist. The incorrigible nature of delusions, for example, has been proposed to reflect increased prior precision at higher hierarchical levels, alongside (and potentially compensating for) weakened prior precision at lower levels [117]. Predictive Coding's focus on prediction is shared by earlier computational theories of psychosis, such as the "aberrant salience" hypothesis and "comparator (corollary discharge) model" [126].

### Relationship to cognitive mapping

Predictive coding and cognitive mapping share a core premise: the brain constructs internal models of the latent causal structure of the environment, a process driven by Bayesian inference. Predictive coding offers a formal process theory for how the brain might approximate Bayesian inference through message passing in cortical circuits. This framework is thus theoretically capable of advancing models linking neurophysiology to brain-wide computational principles [126]. We have suggested that many facets of cognitive mapping might also be fruitfully understood within a similar Bayesian framework, in which structural codes constitute priors for state-space construction and relational inference in new environments.

Despite this convergence, cognitive mapping and predictive coding diverge in their focus. Cognitive mapping emphasizes specific coding motifs in brain regions that support relational inference, goal-directed planning, and task representation. These include abstract structural codes in medial entorhinal cortex, conjunctive state representations in hippocampus, goal-centred map representations in prefrontal cortex, and hippocampal ensemble sequences. Attention is given to how the nature of these coding motifs facilitates efficient knowledge generalization and map construction (e.g., abstraction, factorization) [10,12]. This difference in focus links cognitive mapping theories to a large body of empirical research on spatial map representation in rodents, and domain-general mapping in humans (**Box 3**). It may also cast known neurophysiological disturbances in schizophrenia in a new light. Here, one intriguing question relates to how mesolimbic dopamine signalling and midbrain-hippocampal interaction impacts hippocampal cognitive mapping [69,127,128].

**Box 3: Indexing representations in task-based neuroimaging**

### Decoding and encoding

These are supervised machine learning models that link neural response vectors to task state labels. Decoding predicts labels from neural activity, while encoding predicts neural activity from labels. Their generalization to unseen data (e.g., "leave-one-trial-out" cross-validation) is taken as evidence that neural activity represents task states. In tasks where multiple sensory states map to a single latent state, cross-condition generalization performance (CCGP) can indicate abstract representations (cross-validation leaves out all trials from one sensory state (condition), and labels correspond to abstract states) [48,129].

### Representational geometry

Representational geometry examines how neural responses relate across task states – a second order statistic [130,131]. Repetition suppression uses neural adaptation to quantify neural response relatedness: if A and B share a representation, the neural response to B is reduced after A [47,52,53,63,130]. Representational Similarity Analysis (RSA) quantifies (correlation or Euclidean) distance between each state's multi-voxel/multi-sensor activity vector [41,53,54,84,131] (see **Figure** below). These methods are insensitive to the neural axes of representation - geometry is conserved under rotations.

### Grid coding and replay

Sequential neural replay can be indexed using Temporally Delayed Linear Modelling (TDLM) in MEG, which combines neural decoding and time-lagged regression [47,48,84,132] (**Figure 2C**). Analogous approaches can be used in fMRI, exploiting within- and between-TR variation in decoder performance [59,133]. Grid-like codes can be measured by exploiting the fact that grid cells are organised in modules, where cells within a module share grid spacing (resolution) and orientation (angle). During virtual navigation tasks, module-aligned movement results in increased grid cell firing (a combination of increased grid field sampling, and alignment to conjunctive grid cell preferred firing direction), and this is reflected in neural activity proxies [55–58] (**Figure 2D**).

### Dimensionality reduction

These tools project high-dimensional neural activity to a lower-dimensional space, maximising some objective (e.g., Principal Component Analysis: orthogonal basis set, iteratively maximising explained residual variance). Dimensionality reduction might uncover modes of variation in population activity that mirror the structure of internal representations (cf. neural manifold theories [3,8,129,134]). Several techniques exist – from linear to non-linear; from unsupervised to supervised – mandating careful consideration of the assumptions and biases inherent in any one method [134].

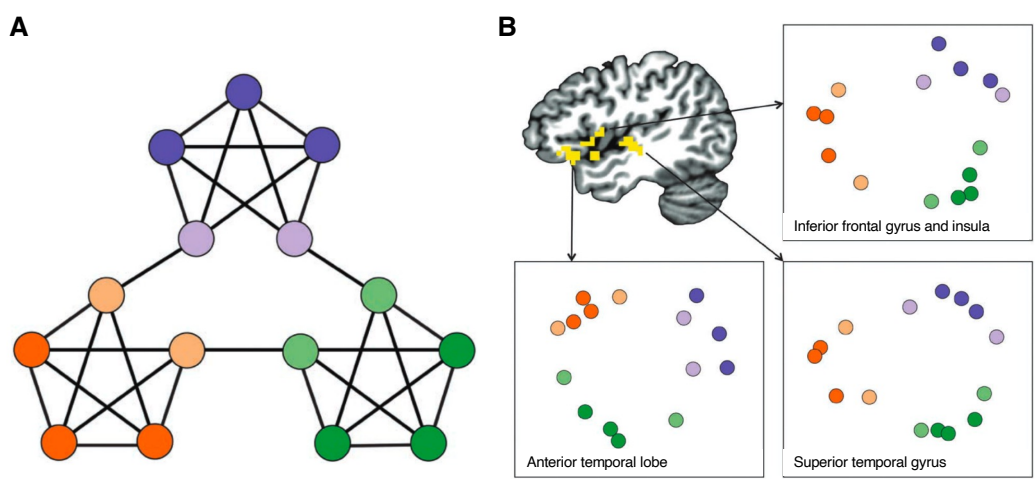

**Representational similarity of non-spatial task.** (A) Task structure used to generate stimulus sequence (each node a picture, each edge a random walk sequential transition) (B) Representational geometry of fMRI responses to each stimulus (node) (searchlight RSA combined with multidimensional scaling). Adapted from [53].

## Acknowledgements

MMN's work is funded by an NIHR Clinical Lectureship in Psychiatry (University of Oxford). YL's work is funded by the Chinese National Science and Technology Innovation 2030 Major Program (2022ZD0205500). ME-G's work is funded by a Wellcome Collaborator award (214314/Z/18/Z). RAM's work is funded by a Wellcome Trust Clinical Research Career Development (224625/Z/21/Z). RJD is supported by the Max Planck Society (MPS). MMN thanks Rick Adams for helpful discussions and comments on an early draft of this manuscript.

## Financial disclosures

RAM has received speaker or consultancy fees from Karuna, Janssen, Boehringer Ingelheim, and Otsuka, and co-directs a company that designs digital resources to support treatment of mental ill health. All other authors declare no competing interests.

## References

1. Craik, K. (1943) *The Nature of Explanation*, Cambridge University Press
2. Andrew Miller and Charles Raison (2023) Burning down the house: reinventing drug discovery in psychiatry for the development of targeted therapies. *Molecular Psychiatry*
3. Nour, M.M. *et al.* (2022) Functional neuroimaging in psychiatry and the case for failing better. *Neuron* 110, 2524–2544
4. Howes, O.D. and Murray, R.M. (2014) Schizophrenia : an integrated sociodevelopmental-cognitive model. *The Lancet* 383, 1677–1687
5. Lee, M. *et al.* (2024) Cognitive Function and Variability in Antipsychotic Drug–Naive Patients With First-Episode Psychosis: A Systematic Review and Meta-Analysis. *JAMA Psychiatry* DOI: 10.1001/jamapsychiatry.2024.0016
6. Knapp, F. *et al.* (2017) Planning performance in schizophrenia patients: a meta-analysis of the influence of task difficulty and clinical and sociodemographic variables. *Psychol. Med.* 47, 2002–2016
7. Kerns, J.G. *et al.* (2008) Executive Functioning Component Mechanisms and Schizophrenia. *Biological Psychiatry* 64, 26–33
8. Barack, D.L. and Krakauer, J.W. (2021) Two views on the cognitive brain. *Nature Reviews Neuroscience* 22, 359–371
9. Bakermans, J.J.W. *et al.* (2024) Constructing future behaviour in the hippocampal formation through composition and replay. *arXiv* DOI: 10.1101/2023.04.07.536053
10. Behrens, T.E.J. *et al.* (2018) What Is a Cognitive Map? Organizing Knowledge for Flexible Behavior. *Neuron* 100, 490–509
11. Bellmund, J.L.S. *et al.* (2018) Navigating cognition: Spatial codes for human thinking. *Science* 362, eaat6766

12. Whittington, J.C.R. *et al.* (2022) How to build a cognitive map. *Nat Neurosci* DOI: 10.1038/s41593-022-01153-y
13. Tolman, E.C. (1948) Cognitive maps in rats and men. *The Psychological Review* 55, 189–208
14. Harlow, H.F. (1949) The formation of learning sets. *Psychological Review* 56, 51–65
15. Piantadosi, S.T. and Gallistel, C.R. (2024) Formalising the role of behaviour in neuroscience. *Eur J of Neuroscience* DOI: 10.1111/ejn.16372
16. Dolan, R.J. and Dayan, P. (2013) Goals and habits in the brain. *Neuron* 80, 312–25
17. Whittington, J.C. *et al.* (2020) The Tolman-Eichenbaum Machine: Unifying space and relational memory through generalisation in the hippocampal formation. *Cell* 183, 1249–1263
18. Tenenbaum, J.B. *et al.* (2011) How to grow a mind: statistics, structure, and abstraction. *Science (New York, N.Y.)* 331, 1279–85
19. Radulescu, A. *et al.* (2021) Human Representation Learning. *Annu. Rev. Neurosci.* 44, 253–273
20. Pezzulo, G. *et al.* (2024) Generating meaning: active inference and the scope and limits of passive AI. *Trends in Cognitive Sciences* 28, 97–112
21. Moser, E.I. *et al.* (2008) Place Cells, Grid Cells, and the Brain's Spatial Representation System. *Annual Review of Neuroscience* 31, 69–89
22. O'Keefe, J. and Dostrovsky, J. (1971) Short Communications The hippocampus as a spatial map . Preliminary evidence from unit activity in the freely-moving rat. *Brain Research* 34, 171–175
23. O'Keefe, J. and Nadel, L. (1978) *The hippocampus as a cognitive map*, Clarendon Press
24. Wilson, M.A. and McNaughton, B.L. (1993) Dynamics of the hippocampal ensemble code for space. *Science* 261, 1055–1058
25. Fyhn, M. *et al.* (2004) Spatial representation in the entorhinal cortex. *Science* 305, 1258–1264
26. Hafting, T. *et al.* (2005) Microstructure of a spatial map in the entorhinal cortex. *Nature* 436, 801–806
27. Høydal, Ø.A. *et al.* (2019) Object-vector coding in the medial entorhinal cortex. *Nature* 568, 400–404
28. Lever, C. *et al.* (2009) Boundary Vector Cells in the Subiculum of the Hippocampal Formation. *J. Neurosci.* 29, 9771–9777
29. Gupta, A.S. *et al.* (2012) Segmentation of spatial experience by hippocampal theta sequences. *Nature Neuroscience* 15, 1032–1039
30. Johnson, A. and Redish, A.D. (2007) Neural Ensembles in CA3 Transiently Encode Paths Forward of the Animal at a Decision Point. *Journal of Neuroscience* 27, 12176–12189
31. Skaggs, W.E. *et al.* (1996) Theta phase precession in hippocampal neuronal populations and the compression of temporal sequences. *Hippocampus* 6, 149–172
32. Wikenheiser, A.M. and Redish, A.D. (2015) Hippocampal theta sequences reflect current goals. *Nature Neuroscience* 18, 289–294

33. Buzsáki, G. (2015) Hippocampal sharp wave-ripple: A cognitive biomarker for episodic memory and planning. *Hippocampus* 25, 1073–1188
34. Foster, D.J. and Wilson, M.A. (2006) Reverse replay of behavioural sequences in hippocampal place cells during the awake state. *Nature* 440, 680–683
35. Lee, A.K. and Wilson, M.A. (2002) Memory of sequential experience in the hippocampus during slow wave sleep. *Neuron* 36, 1183–1194
36. Wilson, M.A. and McNaughton, B.L. (1994) Reactivation of hippocampal ensemble memories during sleep. *Science* 265, 676–679
37. Pfeiffer, B.E. and Foster, D.J. (2013) Hippocampal place-cell sequences depict future paths to remembered goals. *Nature* 497, 74–79
38. Widloski, J. and Foster, D.J. (2022) Flexible rerouting of hippocampal replay sequences around changing barriers in the absence of global place field remapping. *Neuron* 110, 1547-1558.e8
39. Freyja Ólafsdóttir, H. *et al.* (2015) Hippocampal place cells construct reward related sequences through unexplored space. *eLife* 4, 1–17
40. Gupta, A.S. *et al.* (2010) Hippocampal Replay Is Not a Simple Function of Experience. *Neuron* 65, 695–705
41. Barron, H.C. *et al.* (2020) Neuronal Computation Underlying Inferential Reasoning in Humans and Mice. *Cell* 183, 228-243.e21
42. Stella, F. *et al.* (2019) Hippocampal Reactivation of Random Trajectories Resembling Brownian Diffusion. *Neuron* DOI: 10.1016/j.neuron.2019.01.052
43. Dragoi, G. and Tonegawa, S. (2011) Preplay of future place cell sequences by hippocampal cellular assemblies. *Nature* 469, 397–401
44. Kurth-Nelson, Z. *et al.* (2023) Replay and compositional computation. *Neuron* 111, 454–469
45. Dusek, J.A. and Eichenbaum, H. (1997) The hippocampus and memory for orderly stimulus relations. *Proc Natl Acad Sci U S A* 94, 7109–7114
46. Bunsey, M. and Eichenbaum, H. (1996) Conservation of hippocampal memory function in rats and humans. *Nature* 379, 255–257
47. Schwartenbeck, P. *et al.* (2023) Generative replay underlies compositional inference in the hippocampal-prefrontal circuit. *Cell* 186, 4885-4897.e14
48. Liu, Y. *et al.* (2019) Human Replay Spontaneously Reorganizes Experience. *Cell* 178, 640–652
49. Bao, X. *et al.* (2019) Grid-like Neural Representations Support Olfactory Navigation of a Two-Dimensional Odor Space. *Neuron* 102, 1066-1075.e5
50. Aronov, D. *et al.* (2017) Mapping of a non-spatial dimension by the hippocampal-entorhinal circuit. *Nature* 543, 719–722
51. Omer, D.B. *et al.* (2018) Social place-cells in the bat hippocampus. *Science (New York, N.Y.)* 359, 218–224
52. Garvert, M.M. *et al.* (2017) A map of abstract relational knowledge in the human hippocampal–entorhinal cortex. *eLife* 6, 1–20

53. Schapiro, A.C. *et al.* (2013) Neural representations of events arise from temporal community structure. *Nature Neuroscience* 16, 486–492
54. Baram, A.B. *et al.* (2021) Entorhinal and ventromedial prefrontal cortices abstract and generalize the structure of reinforcement learning problems. *Neuron* DOI: 10.1016/j.neuron.2020.11.024
55. Doeller, C.F. *et al.* (2010) Evidence for grid cells in a human memory network. *Nature* 463, 657–661
56. Horner, A.J. *et al.* (2016) Grid-like processing of imagined navigation. *Current Biology* 26, 842–847
57. Constantinescu, A.O. *et al.* (2016) Organizing conceptual knowledge in humans with a gridlike code. *Science* 352, 1464–1468
58. Park, S.A. *et al.* (2021) Inferences on a multidimensional social hierarchy use a grid-like code. *Nature Neuroscience* 24, 1–13
59. Schuck, N.W. and Niv, Y. (2019) Sequential replay of non-spatial task states in the human hippocampus. *Science* 364
60. Kurth-Nelson, Z. *et al.* (2016) Fast Sequences of Non-spatial State Representations in Humans. *Neuron* 91, 194–204
61. Tacikowski, P. *et al.* (2024) Human hippocampal and entorhinal neurons encode the temporal structure of experience. *Nature* DOI: 10.1038/s41586-024-07973-1
62. Samborska, V. *et al.* (2022) Complementary task representations in hippocampus and prefrontal cortex for generalizing the structure of problems. *Nat Neurosci* 25, 1314–1326
63. Barron, H.C. *et al.* (2013) Online evaluation of novel choices by simultaneous representation of multiple memories. *Nature Neuroscience* 16, 1492–1498
64. El-Gaby, M. *et al.* (2023) A Cellular Basis for Mapping Behavioural Structure. *arXiv* DOI: 10.1101/2023.11.04.565609
65. Wilson, R.C. *et al.* (2014) Orbitofrontal Cortex as a Cognitive Map of Task Space. *Neuron* 81, 267–279
66. Schuck, N.W. *et al.* (2016) Human Orbitofrontal Cortex Represents a Cognitive Map of State Space Article Human Orbitofrontal Cortex Represents a Cognitive Map of State Space. *Neuron* 91, 1402–1412
67. Wikenheiser, A.M. *et al.* (2017) Suppression of Ventral Hippocampal Output Impairs Integrated Orbitofrontal Encoding of Task Structure. *Neuron* 95, 1197-1207.e3
68. Zaremba, J.D. *et al.* (2017) Impaired hippocampal place cell dynamics in a mouse model of the 22q11.2 deletion. *Nature Neuroscience* 20, 1612–1623
69. Retailleau, A. and Morris, G. (2018) Spatial Rule Learning and Corresponding CA1 Place Cell Reorientation Depend on Local Dopamine Release. *Current Biology* 28, 836-846.e4
70. Logothetis, N.K. *et al.* (2012) Hippocampal-cortical interaction during periods of subcortical silence. *Nature* 491, 547–553
71. Liu, X. *et al.* (2021) Multimodal neural recordings with Neuro-FITM uncover diverse patterns of cortical–hippocampal interactions. *Nature Neuroscience* 24, 886–896

72. Higgins, C. *et al.* (2021) Replay bursts in humans coincide with activation of the default mode and parietal alpha networks. *Neuron* 109, 882–893
73. Kaplan, R. *et al.* (2016) Hippocampal Sharp-Wave Ripples Influence Selective Activation of the Default Mode Network. *Current Biology* 26, 686–691
74. Huang, Q. *et al.* (2024) Replay-triggered brain-wide activation in humans. *Nat Commun* 15, 7185
75. Yeshurun, Y. *et al.* (2021) The default mode network: where the idiosyncratic self meets the shared social world. *Nature Reviews Neuroscience* 22, 181–192
76. Margulies, D.S. *et al.* (2016) Situating the default-mode network along a principal gradient of macroscale cortical organization. *Proc. Natl. Acad. Sci. U.S.A.* 113, 12574–12579
77. Hahamy, A. *et al.* (2023) The human brain reactivates context-specific past information at event boundaries of naturalistic experiences. *Nat Neurosci* 26, 1080–1089
78. Baldassano, C. *et al.* (2017) Discovering Event Structure in Continuous Narrative Perception and Memory. *Neuron* 95, 709-721.e5
79. Morris, R.W. *et al.* (2018) Impairments in action-outcome learning in schizophrenia. *Translational Psychiatry* 8
80. Culbreth, A.J. *et al.* (2016) Reduced model-based decision-making in schizophrenia. *J Abnorm Psychol.* 125, 777–787
81. Titone, D. *et al.* (2004) Transitive inference in schizophrenia: Impairments in relational memory organization. *Schizophrenia Research* 68, 235–247
82. Adams, R.A. *et al.* (2020) Impaired theta phase coupling underlies frontotemporal dysconnectivity in schizophrenia. *Brain* 143, 1261–1277
83. Armstrong, K. *et al.* (2012) Impaired associative inference in patients with schizophrenia. *Schizophrenia Bulletin* 38, 622–629
84. Nour, M.M. *et al.* (2021) Impaired neural replay of inferred relationships in schizophrenia. *Cell* 184
85. Thai, M.L. *et al.* (2019) A meta-analysis of executive dysfunction in patients with schizophrenia: Different degree of impairment in the ecological subdomains of the Behavioural Assessment of the Dysexecutive Syndrome. *Psychiatry Research* 272, 230–236
86. Fradkin, I. and Eldar, E. (2023) Accumulating evidence for myriad alternatives: Modeling the generation of free association. *Psychological Review* 130, 1492–1520
87. Lundin, N.B. *et al.* (2023) Neural evidence of switch processes during semantic and phonetic foraging in human memory. *Proc. Natl. Acad. Sci. U.S.A.* 120, e2312462120
88. Hills, T.T. *et al.* (2012) Optimal foraging in semantic memory. *Psychological Review* 119, 431–440
89. Solomon, E.A. *et al.* (2019) Hippocampal theta codes for distances in semantic and temporal spaces. *Proceedings of the National Academy of Sciences* DOI: 10.1101/611681

90. Corcoran, C.M. and Cecchi, G.A. (2020) Using Language Processing and Speech Analysis for the Identification of Psychosis and Other Disorders. *Biological Psychiatry: Cognitive Neuroscience and Neuroimaging* 5, 770–779

91. Nour, M. *et al.* (2023) Trajectories through semantic spaces in schizophrenia and the relationship to ripple bursts. *Proceedings of the National Academy of Sciences (PNAS)* 120

92. Suh, J. *et al.* (2013) Impaired Hippocampal Ripple-Associated Replay in a Mouse Model of Schizophrenia. *Neuron* 80, 484–493

93. Altimus, C. *et al.* (2015) Disordered Ripples Are a Common Feature of Genetically Distinct Mouse Models Relevant to Schizophrenia. *Molecular Neuropsychiatry* 1, 52–59

94. Nour, M.M. *et al.* (2023) Reduced coupling between offline neural replay events and default mode network activation in schizophrenia. *Brain Communications* 5, fcad056

95. Nour, M. *et al.* (2022) Relationship between replay-associated ripple power and hippocampal NMDA receptor binding in schizophrenia. *Schizophrenia Bulletin Open*

96. Dupret, D. *et al.* (2010) The reorganization and reactivation of hippocampal maps predict spatial memory performance. *Nature Neuroscience* 13, 995–1002

97. Krystal, J.H. *et al.* (2017) Impaired Tuning of Neural Ensembles and the Pathophysiology of Schizophrenia: A Translational and Computational Neuroscience Perspective. *Biological Psychiatry* 81, 874–885

98. Rolls, E.T. *et al.* (2008) Computational models of schizophrenia and dopamine modulation in the prefrontal cortex. *Nat Rev Neurosci* 9, 696–709

99. Sigurdsson, T. *et al.* (2010) Impaired hippocampal–prefrontal synchrony in a genetic mouse model of schizophrenia. *Nature* 464, 763–767

100. Dickerson, D.D. *et al.* (2010) Abnormal Long-Range Neural Synchrony in a Maternal Immune Activation Animal Model of Schizophrenia. *J. Neurosci.* 30, 12424–12431

101. Convertino, L. *et al.* (2022) Reduced grid-like theta modulation in schizophrenia. *Brain* 146, 2191–2198

102. Brugger, S.P. and Howes, O.D. (2017) Heterogeneity and Homogeneity of Regional Brain Structure in Schizophrenia A Meta-analysis. *JAMA psychiatry* 74, 1104–1111

103. Radhakrishnan, R. *et al.* (2021) In vivo evidence of lower synaptic vesicle density in schizophrenia. *Mol Psychiatry* 26, 7690–7698

104. McHugo, M. *et al.* (2019) Hyperactivity and Reduced Activation of Anterior Hippocampus in Early Psychosis. *American Journal of Psychiatry* DOI: 10.1176/appi.ajp.2019.19020151

105. Howes, O.D. and Onwordi, E.C. (2023) The synaptic hypothesis of schizophrenia version III: a master mechanism. *Mol Psychiatry* 28, 1843–1856

106. Onwordi, E.C. *et al.* (2020) Synaptic density marker SV2A is reduced in schizophrenia patients and unaffected by antipsychotics in rats. *Nature Communications* 11

107. Howes, O.D. and Shatalina, E. (2022) Integrating the Neurodevelopmental and Dopamine Hypotheses of Schizophrenia and the Role of Cortical Excitation-Inhibition Balance. *Biological Psychiatry* 92, 501–513

108. Hamm, J.P. *et al.* (2017) Altered Cortical Ensembles in Mouse Models of Schizophrenia. *Neuron* 94, 153-167.e8
109. Hamm, J.P. *et al.* (2020) Aberrant Cortical Ensembles and Schizophrenia-like Sensory Phenotypes in Setd1a+/− Mice. *Biological Psychiatry* 88, 215–223
110. Musa, A. *et al.* (2022) The Shallow Cognitive Map Hypothesis : A Hippocampal framework for Thought Disorder in Schizophrenia. *Schizophrenia* 8
111. Hauser, T.U. *et al.* (2016) Computational Psychiatry of ADHD: Neural Gain Impairments across Marrian Levels of Analysis. *Trends in Neurosciences* 39, 63–73
112. Adams, R.A. *et al.* (2018) Attractor-like dynamics in belief updating in schizophrenia. *The Journal of Neuroscience* DOI: 10.1523/JNEUROSCI.3163-17.2018
113. Munn, R.G.K. *et al.* (2023) Disrupted hippocampal synchrony following maternal immune activation in a rat model. *Hippocampus* 33, 995–1008
114. Winship, I.R. *et al.* (2019) An Overview of Animal Models Related to Schizophrenia. *Can J Psychiatry* 64, 5–17
115. Speers, L.J. *et al.* (2021) Hippocampal Sequencing Mechanisms Are Disrupted in a Maternal Immune Activation Model of Schizophrenia Risk. *J. Neurosci.* 41, 6954–6965
116. Mesbah-Oskui, L. *et al.* (2015) Hippocampal place cell and inhibitory neuron activity in disrupted-in-schizophrenia-1 mutant mice: implications for working memory deficits. *npj Schizophrenia* 1, 1–7
117. Petrovic, P. and Sterzer, P. (2023) Resolving the Delusion Paradox. *Schizophrenia Bulletin* 49, 1425–1436
118. Zhao, B. *et al.* (2023) A model of conceptual bootstrapping in human cognition. *Nat Hum Behav* 8, 125–136
119. Hofstadter, D. (2001) Analogy as the Core of Cognition. In *The Analogical Mind: Perspectives from Cognitive Science*, pp. 499–538, MIT Press
120. Nussenbaum, K. and Hartley, C.A. (2024) Understanding the development of reward learning through the lens of meta-learning. *Nat Rev Psychol* DOI: 10.1038/s44159-024-00304-1
121. Whittington, J.C.R. *et al.* (2023) Disentangling with Biological Constraints: A Theory of Functional Cell Types. *ICLR* at <http://arxiv.org/abs/2210.01768>
122. Doerig, A. *et al.* (2023) The neuroconnectionist research programme. *Nat Rev Neurosci* 24, 431–450
123. Kay, S. *et al.* (1987) The Positive and Negative Syndrome Scale (PANSS) for schizophrenia. *Schizophr Bull.* 13, 261–276
124. Hall, A.F. *et al.* (2024) The computational structure of consummatory anhedonia. *Trends in Cognitive Sciences* DOI: 10.1016/j.tics.2024.01.006
125. Ritunnano, R. *et al.* (2022) Subjective experience and meaning of delusions in psychosis : a systematic review and qualitative evidence synthesis. *The Lancet Psychiatry* 9, 458–476
126. Sterzer, P. *et al.* (2018) The Predictive Coding Account of Psychosis. *Biological Psychiatry* 84, 634–643

127. Gomperts, S.N. *et al.* (2015) VTA neurons coordinate with the hippocampal reactivation of spatial experience. *eLife* 4, 1–22
128. McNamara, C.G. *et al.* (2014) Dopaminergic neurons promote hippocampal reactivation and spatial memory persistence. *Nature Neuroscience* 17, 1658–1660
129. Bernardi, S. *et al.* (2020) The Geometry of Abstraction in the Hippocampus and Prefrontal Cortex. *Cell* DOI: 10.1016/j.cell.2020.09.031
130. Barron, H.C. *et al.* (2016) Repetition suppression: a means to index neural representations using BOLD? *Philosophical Transactions of the Royal Society B: Biological Sciences* 371, 20150355
131. Diedrichsen, J. and Kriegeskorte, N. (2017) Representational models: A common framework for understanding encoding, pattern-component, and representational-similarity analysis. *PLoS Computational Biology* 13, 1–33
132. Liu, Y. *et al.* (2021) Temporally delayed linear modelling (TDLM) measures replay in both animals and humans. *eLife* 10, e66917
133. Wittkuhn, L. and Schuck, N.W. (2021) Dynamics of fMRI patterns reflect sub-second activation sequences and reveal replay in human visual cortex. *Nature Communications* 12
134. Roads, B.D. and Love, B.C. (2024) The Dimensions of dimensionality. *Trends in Cognitive Sciences* DOI: 10.1016/j.tics.2024.07.005